\documentclass[10pt,superscriptaddress,twocolumn,amsmath,amssymb]{revtex4-1}

\usepackage{graphicx} 
\usepackage[colorlinks=true,linkcolor=blue,citecolor=blue,urlcolor=black]{hyperref}
\usepackage[english]{babel}

\usepackage{datetime}
\usepackage{verbatim}
\usepackage{amsmath}
\usepackage{amssymb}
\usepackage{siunitx} 
\usepackage{nccmath}
\usepackage{dcolumn}
\usepackage{lineno}
\usepackage{xcolor}
\usepackage{color}
\usepackage{float}

\usepackage[utf8]{inputenc}
\usepackage{lineno}

\addto\captionsenglish{}

\definecolor{green}{rgb}{0,0.5,0}

\begin{document}

\widetext

\title{Activity-induced microswimmer interactions and cooperation in one-dimensional environments}

\author{Stefania Ketzetzi}
\affiliation{Soft Matter Physics, Huygens-Kamerlingh Onnes Laboratory, Leiden University, P.O. Box 9504, 2300 RA Leiden, The Netherlands}
\author{Melissa Rinaldin}
\affiliation{Soft Matter Physics, Huygens-Kamerlingh Onnes Laboratory, Leiden University, P.O. Box 9504, 2300 RA Leiden, The Netherlands}
\affiliation{Current address: Max Planck Institute of Molecular Cell Biology and Genetics, Pfotenhauerstr. 108, 01307 Dresden}
\author{Pim Dr\"oge}
\affiliation{Soft Matter Physics, Huygens-Kamerlingh Onnes Laboratory, Leiden University, P.O. Box 9504, 2300 RA Leiden, The Netherlands}
\author{Joost de Graaf}
\affiliation{Institute for Theoretical Physics, Center for Extreme Matter and Emergent Phenomena, Utrecht University, Princetonplein 5, 3584 CC Utrecht, The Netherlands}
\author{Daniela J. Kraft}
\email{Corresponding author: kraft@physics.leidenuniv.nl}
\affiliation{Soft Matter Physics, Huygens-Kamerlingh Onnes Laboratory, Leiden University, P.O. Box 9504, 2300 RA Leiden, The Netherlands}

\date{\today}

\begin{abstract}
Cooperative motion in biological microswimmers is crucial for their survival as it facilitates adhesion to surfaces, formation of hierarchical colonies, efficient motion, and enhanced access to nutrients. Synthetic microswimmers currently lack truly cooperative behavior that originates from activity-induced interactions. Here, we demonstrate that catalytic microswimmers show a variety of cooperative behaviors along one-dimensional paths. We show that their speed increases with the number of swimmers, while the activity induces a preferred distance between swimmers. Using a minimal model, we ascribe this behavior to an effective activity-induced potential that stems from a competition between chemical and hydrodynamic coupling. These interactions further induce active self-assembly into trains as well as compact chains that can elongate, break-up, become immobilized and remobilized. We identify the crucial role that environment morphology and swimmer directionality play on these highly dynamic chain behaviors. These activity-induced interactions open the door towards exploiting cooperation for increasing the efficiency of, as well as provide temporal and spatial control over, microswimmer motion, thereby enabling them to perform intricate tasks inside complex environments.
\end{abstract}

\maketitle

Many microorganisms crucially rely on cooperation for their survival and thriving. Cooperation greatly enhances microorganism motility and overall motion efficiency~\cite{Elgeti2015}, and often leads to the formation of organized, complex colonies. For example, spermatozoa self-assemble into train-like structures to enhance fertilization~\cite{Moore2002}, \textit{Volvox} algae form colonies to propel and facilitate fluid flows with nutrients and chemical messengers~\cite{Solari2007}, and cancer cells secrete chemicals to communicate and promote tumour growth~\cite{Archetti2019}. Similarly, bacteria cooperate to enhance surface adhesion~\cite{Lee2020} during biofilm formation which increases their resistance to environmental stresses and drugs, their spreading, and the efficiency of nutrient capture~\cite{Lee2020, Santos2018, Stoodley2004}. At high densities, bacterial colonies again rely on cooperation to form swarms with large-scale dynamic patterns, such as whirls and jets, to expand and to explore their surroundings while simultaneously reducing their competition for nutrients~\cite{Beer2019,Beer2020}. These vital behaviors are achieved by exploiting interactions based on hydrodynamic and steric effects~\cite{Drescher2011}, as well as chemical signaling, called quorum-sensing~\cite{Miller2001}.

Similar to their biological counterparts, synthetic swimmers are also able to perform directed motion inside liquid environments~\cite{Katuri2017}, even under realistic conditions such as inside patterned~\cite{Volpe2011, Takagi2014, Das2015, Simcchen2015, Yu2016, Liu2016, Brown2016,  Wykes2017, Katuri2018,Mijalkov2013,Jin2020} and biological environments~\cite{Wang2014,Schamel2014,Walker2015}. Consequently, they offer exciting opportunities as they open new routes for precise motion control, and hence a wide range of applications, on the microscale~\cite{Katuri2017}. In the future, these swimmers may be deployed to perform complicated tasks such as \textit{in vivo} drug delivery and repair~\cite{Patra2013, Gao2015} inside complex and crowded environments~\cite{Bechinger2016, Ebbens2016}, e.g. in living organisms and lab-on-a-chip devices~\cite{Garcia2013, Restrepo2014}.

Drawing inspiration from biological systems and their efficiency-increasing strategies, it is desirable that the aforementioned tasks are performed not only on the single but on the multi-swimmer level~\cite{Gao2013, Ni2017}. For instance, if employed in drug delivery, collections of swimmers may reach the desired target faster, or deliver a higher dosage~\cite{Cheang2014}. Cooperative behavior and communication between the microswimmers could furthermore enable different types of delivery, in which for example dosages are applied at specific times or time intervals~\cite{Tsang2020}.

Although \textit{collective} effects, such as enhanced aggregation, cluster and crystal formation, ordering and phase separation, have been observed for synthetic systems in 2- and 3-dimensions~\cite{Theurkauff2012,Buttinoni2013,Palacci2013,Wysocki2014}, these effects can in principle be explained by volume exclusion and persistent motion of the swimmer~\cite{Buttinoni2013,Wysocki2014,Elgeti2015}, and do not require \textit{cooperation}; with the latter typically relying on information exchange to enhance the efficiency of their behavior. Even the exciting recently observed corralling of passive particles by swarms of light-driven synthetic swimmers was explained purely by geometric arguments~\cite{Zion2021}, and the formation of self-spinning microgears~\cite{Aubret2018} and active colloidal molecules~\cite{Wang2020} required external fields for their assembly and/or propulsion. While cooperation is a type of collective effect, the inverse is not true. Thus, the collective behavior of synthetic microswimmers at higher densities does not necessarily signify that they collaborate and cooperate in the same sense as biological swimmers, which employ signaling and sensing, do. 

\begin{figure*}[!ht]
    \centering
     \includegraphics[width=1\textwidth]{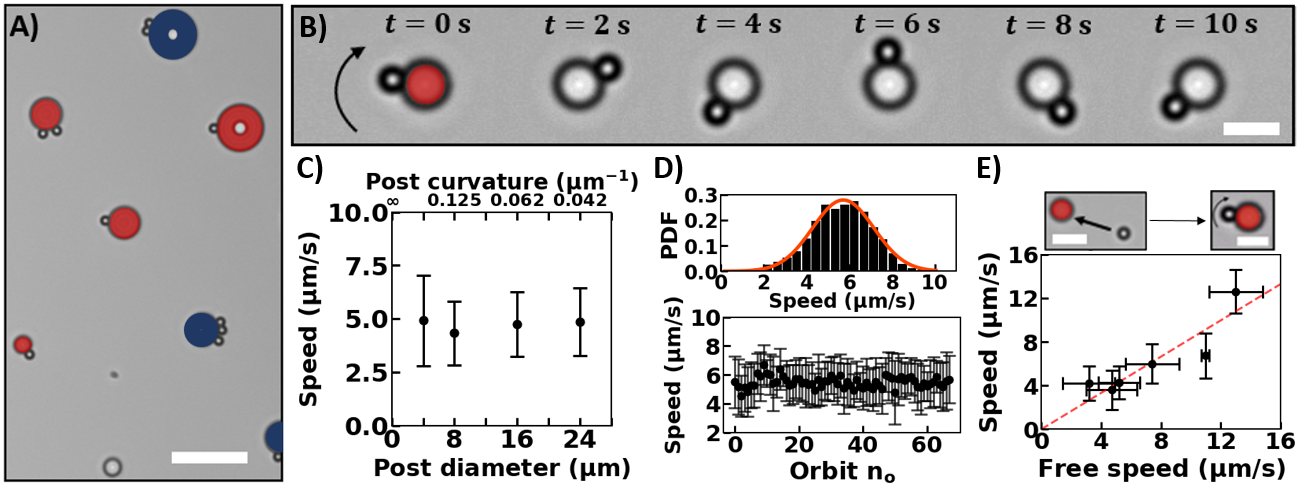}
    \caption{\textbf{Catalytic microswimmer motion along circular posts: Single particle orbiting.} 
    \textbf{A)} Light microscopy image showing our catalytic microswimmers in one of our experimental setups featuring circular 3D micro-printed posts on planar substrates. Coloring is used to distinguish the stationary posts from our swimmers, and indicates whether the attached swimmers were mobile (red posts) or immobile (blue posts). Scale bar is 20~{\textmu}m.
    \textbf{(B)} Time series of light microscopy images of a (2.00 $\pm$ 0.05)~{\textmu}m diameter swimmer orbiting a 4~{\textmu}m diameter post, with the arrow denoting its constant direction of motion. Scale bar is 5~{\textmu}m. \textbf{(C)} Bottom: propulsion speed along the post as function of post diameter. All data were taken at otherwise fixed experimental conditions. \textbf{(D)} Propulsion speed as function of number of orbits of particle shown in (B). Measurement duration was 4 min. Top: same data plotted as a probability density, showing that speed in orbit follows a Gaussian distribution. \textbf{(E)} Propulsion speed for the same swimmers before orbiting, i.e. ``Free speed" on the planar substrate, plotted against its speed in orbit. The dashed line is a least-squares fit with $y = ax$ and $a = 0.83 \pm 0.08$, in line with Ref.~\cite{Takagi2014}.}
    \label{fig:fig1}
\end{figure*} 
Here, we demonstrate that catalytically propelled model microswimmers exhibit a wealth of activity-induced interactions along closed 1-dimensional paths. Single swimmers move at stable speeds independent of path curvature. We furthermore find that multiple swimmers along the same path exhibit cooperation. That is, their speed increases with the number of swimmers while their activity induces a preferred distance between them. We show that swimmer cooperation originates from a combination of steric effects and hydrodynamic and chemical couplings. Lastly, we demonstrate rich locomotion behavior induced by chain fission and fusion, which has only been considered theoretically in the context of magnetic swimmers~\cite{Lastra2016}. We reveal that chain formation and breakup can be tuned using the local curvature of the path.

\section*{Results}
To study their dynamic behavior and interactions, we confine swimmers to 1-dimensional tracks by exploiting their strong affinity for surfaces~\cite{Das2015,Simcchen2015,Brown2016, Ketzetzi2020, Ketzetzi2020sep}. This affinity stems from their propulsion mechanism, which is based on an asymmetric catalytic decomposition of H$_2$O$_2$ on their Pt-coated hemisphere~\cite{Golestanian2005}. We equip planar substrates with designed 3D micro-printed posts, thereby effectively creating preferred 1-dimensional environments around the posts, where the swimmers can be in close vicinity to both posts and substrates~\cite{Das2015,Simcchen2015,Brown2016}. See Figure~\ref{fig:fig1}A for one example of our experimental setups featuring circular posts. The H$_2$O$_2$ decomposition reaction here creates gradients in solute molecules which act over the swimmer surfaces, posts and substrates, inducing phoretic and osmotic flows thereby causing swimmer capture~\cite{Ketzetzi2020, Ketzetzi2020sep}. That is, when a swimmer encounters a post, it quickly gets captured into motion along it, and is retained there for very long times. 

Once attached to a circular post, our swimmers with diameters of (2.00 $\pm$ 0.05)~{\textmu}m and Pt coating thicknesses of (4.7 $\pm$ 0.2) nm moved with equal probability in either the clockwise or counterclockwise direction, without switching direction. Single swimmers, as well as multiple swimmers with the same direction of motion, on a given post, orbited their posts with approximately constant speed. We highlighted these swimmers by coloring their corresponding posts in red in Figure~\ref{fig:fig1}A. These swimmers continued their orbiting motion for at least 30 min in our experiments, during which we did not see them leave. However, swimmers that moved in opposing directions on a given post, hindered each others motion and were immobilized, see Supplementary Movie 1 as an example. We indicated these immobilized clusters by coloring their adjacent posts blue in Figure~\ref{fig:fig1}A. 

\begin{figure*}[ht]
    \centering
    \includegraphics[width=1\linewidth]{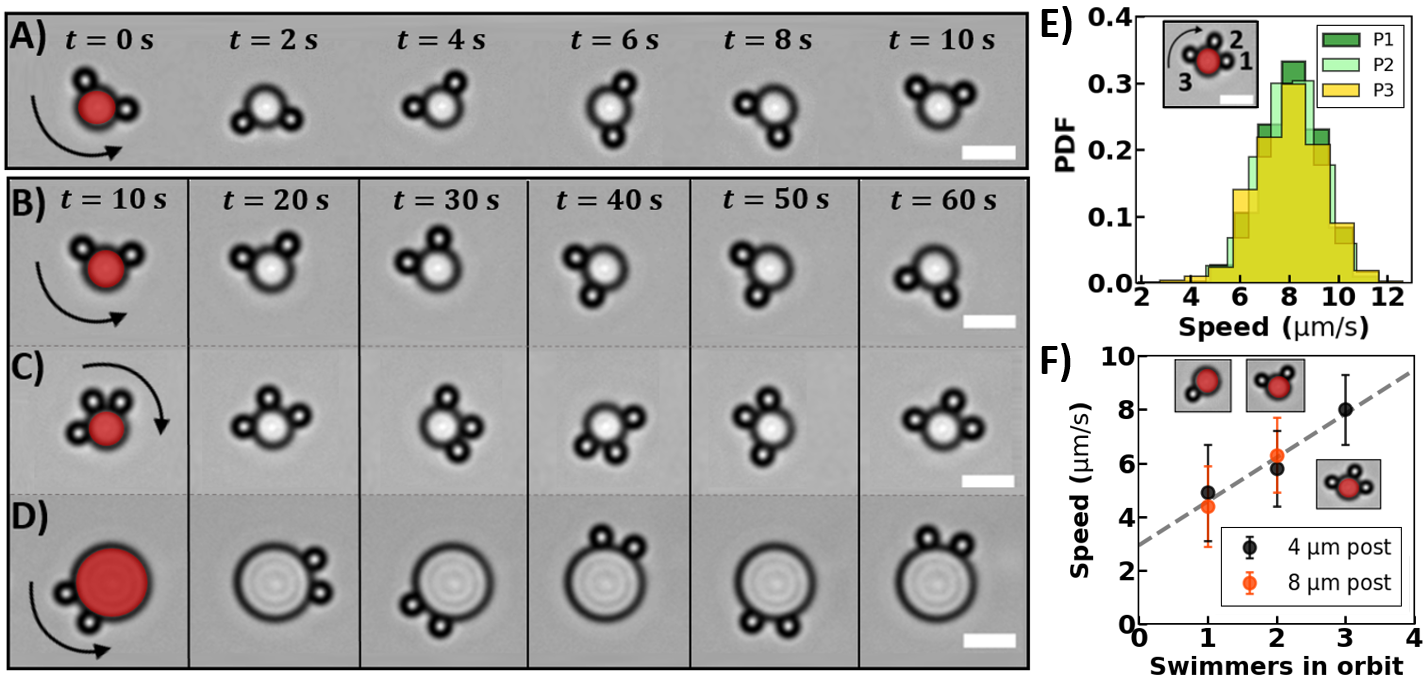}
    \caption{\textbf{Multiple microswimmers orbiting circular posts.} Red color has been added to distinguish the stationary posts from the orbiting swimmers. Scale bars are 5~{\textmu}m. Sequence of light microscopy images showing orbiting of \textbf{A, B)} two and \textbf{C)} three swimmers around a 4~{\textmu}m diameter post, and \textbf{D)}~two swimmers around an 8~{\textmu}m diameter post. See also Supplementary Movies 2-4. Each column in panels (B)-(D) shares the same timestamp, indicated in (B). \textbf{(E)} The PDFs of propulsion speed in orbit for three swimmers on a post are almost identical. \textbf{(F)} Average propulsion speed as a function of the number of swimmers in orbit for two post sizes, showing that speed increases with the number of comoving swimmers for both post sizes in a similar fashion. All measurements were taken under the same experimental conditions. The dashed line is a least squares fit with $y=\alpha x +b$, with $\alpha$ and $b$ being (1.6 $\pm$ 0.2)~{\textmu}m/s and (2.9 $\pm$ 0.4)~{\textmu}m/s, respectively.}
    \label{fig:fig2}
\end{figure*}

Surprisingly, long-term capture around posts happens even when post size is comparable to the swimmer size, see for example Figure~\ref{fig:fig1}B for a swimmer orbiting a 4~{\textmu}m diameter post. This is in stark contrast with simulations on model squirmers~\cite{Spagnolie2015, Kuron2019}, which are often used to approximate synthetic swimmers such as ours. For model squirmers, capture has only been proposed for posts of sizes several times larger than the squirmer~\cite{Spagnolie2015}. Moreover, simulations for the orbiting of a squirmer around a sphere predict a relation between curvature and squirmer speed, as can be readily obtained from the data in Ref.~\cite{Kuron2019}. To test these predictions, we track the swimmers in time and extract their speed in orbit using python routines~\cite{Trackpy}. We find that post curvature does not influence the propulsion speeds of our catalytic swimmers, at least not for the range shown in Figure~\ref{fig:fig1}C. These discrepancies between our experiments and simulations imply that catalytic swimmers are different from pure squirmers, probably due to their propulsion mechanism and to the long-range solute gradients that act across the substrate and posts~\cite{Ketzetzi2020,Ketzetzi2020sep}. 

Despite orbiting short tracks, we find that propulsion speed in orbit is rather stable, see Figure~\ref{fig:fig1}D which shows that speed in orbit follows a Gaussian distribution. Note that the presence of the post itself does not have a considerable effect on propulsion speed: the speed of a swimmer in orbit is only slightly reduced with respect to its ``free" speed parallel to the substrate alone in Figure~\ref{fig:fig1}E. The dashed line is a least-squares fit with $y = ax$ where $a$ is ($0.83\pm0.08$), in line with the slope obtained previously for bimetallic microrods~\cite{Takagi2014}. These findings point out that any hydrodynamic and/or phoretic coupling to the post leads to a subdominant contribution to the speed.

Intriguing effects take place when multiple swimmers orbit the same post: the swimmers move with seemingly equal speeds while also maintaining roughly constant distances, see Figure~\ref{fig:fig2} and Supplementary Movies 2-4. This constancy in speed and distance appears both for two (Figure ~\ref{fig:fig2}A, B) and three comoving swimmers (Figure~\ref{fig:fig2}C), and is independent of the post curvature (Figure~\ref{fig:fig2}D). More quantitatively, we find that swimmers that orbit around the same post in fact have almost the same speed distribution, independent of the particle number and post size. See Figure~\ref{fig:fig2}E for the speed distributions of three swimmers on a 4~{\textmu}m post and Supplemental Information (SI) for additional data. 

Even more strikingly, speed increases with the number of swimmers in orbit, as shown in Figure~\ref{fig:fig2}F for posts with 4 and 8~{\textmu}m diameters. That is, two particles orbit faster than one, and in turn three particles orbit faster than two: Under otherwise fixed conditions, speed increases by $\approx$ 20\% and 60\% for two and three comoving swimmers on 4~{\textmu}m posts, respectively, and by $\approx$ 40\% for two comoving swimmers on the 8~{\textmu}m post, in comparison to single swimmers. Interestingly, for our system there is no significant speedup of a pair of particles with respect to a separated third particle moving along the post. This contrasts strongly with the result of passive, driven particles in a torroidal optical trap~\cite{Curtis2003,Reichert2004,Lutz2006}, where pair formation and breaking is observed. That is, two driven particles overtake a third, which then leads to a fracture of the triplet with the two lead particles moving off, a result that can be understood using hydrodynamic theory~\cite{Curtis2003,Reichert2004,Lutz2006}.

These findings strongly suggest \textit{cooperative motion} of the microswimmers. Chemical gradients of swimmers in close proximity interact, thereby causing multi-bound states and a collective speedup. This speedup is independent of the post size for the here considered 4 and 8~{\textmu}m posts, see Figure~\ref{fig:fig2}F. The dashed line represents a least squares fit with $y = \alpha x+b$, with $\alpha$ (1.6 $\pm$ 0.2)~{\textmu}m/s and $b$ (2.9 $\pm$ 0.4)~{\textmu}m/s, implying a linear relationship between the number of swimmers and the collective speedup. This observation of a speedup as well as the seemingly constant swimmer distance in Figure~\ref{fig:fig2}, is surprising in view of the distribution in speed that single swimmers exhibit, and implies that swimmers experience a coupling that adjusts their speed. To understand the origin of this behavior, we will now quantify the swimmer distances as well as test for correlations of the speeds of pairs of comoving swimmers. 

\begin{figure*}[t]
    \centering
     \includegraphics[width=1\linewidth]{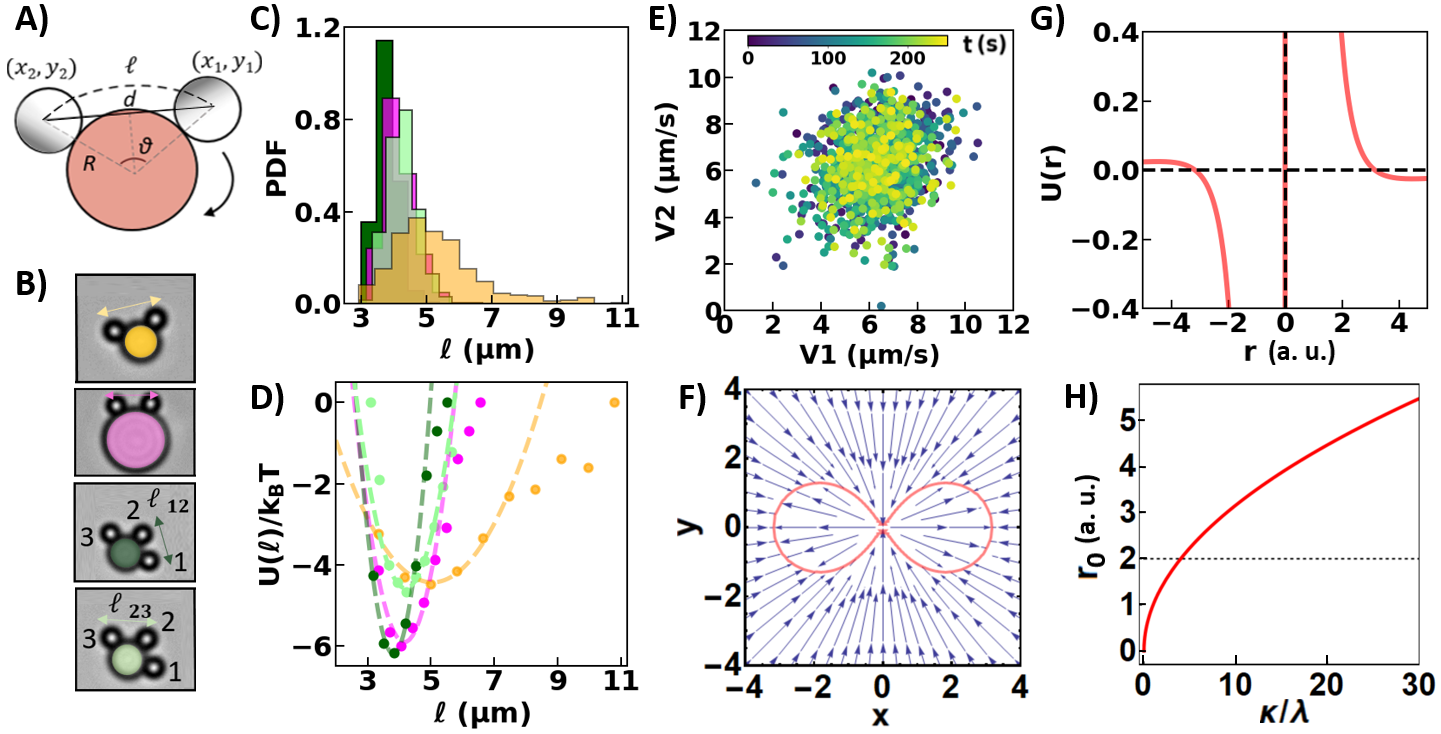}
    \caption{\textbf{Interactions of microswimmers comoving along circular posts: Experiments and modeling.} \textbf{A)} Representation of the arc distance. \textbf{B)} Snapshots of swimmer pairs in orbit, color-coded so that each color marks the corresponding swimmer pair in panels (C-D). \textbf{C)} PDF of the arc distance between comoving swimmers, showing that swimmers assume a relatively constant distance in orbit with minimum center-to-center distance $\approx$ 3~{\textmu}m. Measurement duration is $\approx$ 5 min. \textbf{(D)} Potential energy in terms of thermal energy, as obtained from the swimmer distances in (C) using the Boltzmann distribution. The dashed lines represent least-squares fits with $y = \frac{1}{2}k(x-x_0)^2+y_0$. All fitted parameters are listed in the SI. \textbf{E)} Scatter plot of the speeds of the two swimmers comoving along the 8~{\textmu}m post colored magenta in panels (B-D), showing that swimmer speeds are not correlated. \textbf{(F-H)} Effective separation between swimmers based on hydrodynamic and osmotic balance. \textbf{(F)} The balance between an inward osmotic flow along the wall and an outward pusher-type dipolar flow away from the swimmer leads to a curve of zero velocity depicted in red. \textbf{(G)} Two swimmers that lie head-to-tail as in panel (A) assume a fixed distance as evidenced by the x-axis intercept. The shape of the relative velocity generates an effective potential. \textbf{(H)} Swimmer separation distance as a function of the ratio between the osmotic and pusher contributions, with $\lambda$ and $\kappa$ indicating the respective strengths of the inward and outward flows.}
    \label{fig:fig3}
\end{figure*}

We quantify the arc distance, $\ell$, between comoving swimmers as depicted in Figure~\ref{fig:fig3}A. We measure the distances between various swimmer pairs, on differently sized posts and with different number of attached swimmers, see Figure~\ref{fig:fig3}B. In all cases, we find that swimmers orbit the posts at a preferred distance (Figure~\ref{fig:fig3}C), in line with our expectations based on Figure~\ref{fig:fig2}. Specifically, the average distance measured over a 4~min time interval is (5.3 $\pm$ 1.4)~{\textmu}m and (4.2 $\pm$ 0.5)~{\textmu}m for two swimmers on a 4 and 8~{\textmu}m post, respectively, and (3.8 $\pm$ 0.3)~{\textmu}m and (4.3 $\pm$ 0.4)~{\textmu}m between the front-middle and middle-back swimmers of a three-swimmer system on a 4~{\textmu}m post, respectively. We notice that swimmers never approach closer than a minimum center-to-center distance of (3.0 $\pm$ 0.1)~{\textmu}m. In addition, the arc distances we find here show similar distributions, albeit with different peak values and widths. 

These common features raise the question whether the constancy in the distance stems from correlations between the speeds of the comoving swimmers. However, a closer examination reveals that speeds always vary independently of one another, see the scatter plot of the speeds of two comoving swimmers along an 8~{\textmu}m post, V$_1(t)$ vs V$_2(t)$, where $t$ is time in Figure~\ref{fig:fig3}E and SI. Already, the symmetric shape of the scatter plot indicates that they are not correlated, which is further supported by a Pearson correlation coefficient of 0.2. In addition, we excluded time-delayed correlations by considering the correlation between $V_1(t)$ and $V_2(t+\tau$) see SI, with $\tau$ the time between two frames. Again, the Pearson correlation coefficient was 0.1, signifying no speed correlation. Hence, there must be an alternative mechanism that can explain why the swimmers move with the same average speed and stable distance. 

The random fluctuations around an average distance is reminiscent of the thermal motion of a passive particle inside a potential well. Here, this effective potential must be induced by the active state of \textit{both} swimmers. A single active particle is not sufficient, because passive particles in the vicinity of an active particle do not get confined but are instead either dragged along the fluid flow around the active particle~\cite{Campbell2019} or attracted to the active particle site~\cite{Palacci2013}. 

Seeing that our swimmers are moving together at the same average speed, we consider their activity-confined motion in the rotating frame that comoves with the swimmers. While this is a non-inertial frame of reference and an out-of-equilibrium state, it still allows us to view our system from a simplified perspective: that of random thermal motion of each particle in a potential induced by the other. To do so, we use the Boltzmann distribution, $\mathrm{PDF}(\ell) = \exp{(-U(\ell)/k_BT)}$ to extract the relative potential energy for our systems from the probability density function of their corresponding distances. This allows us to measure the probability for the system to be in a certain distance relative to the distance's energy and system temperature in Figure~\ref{fig:fig3}D, where the relative energy is $U(\ell)/k_BT = -\log (\mathrm{PDF}(\ell)) + U/k_BT$, with $U/k_BT$ an arbitrary reference state, set such that the relative energy goes to 0 at infinite distances. 

\begin{figure*}[ht!]
    \centering
     \includegraphics[width=1\linewidth]{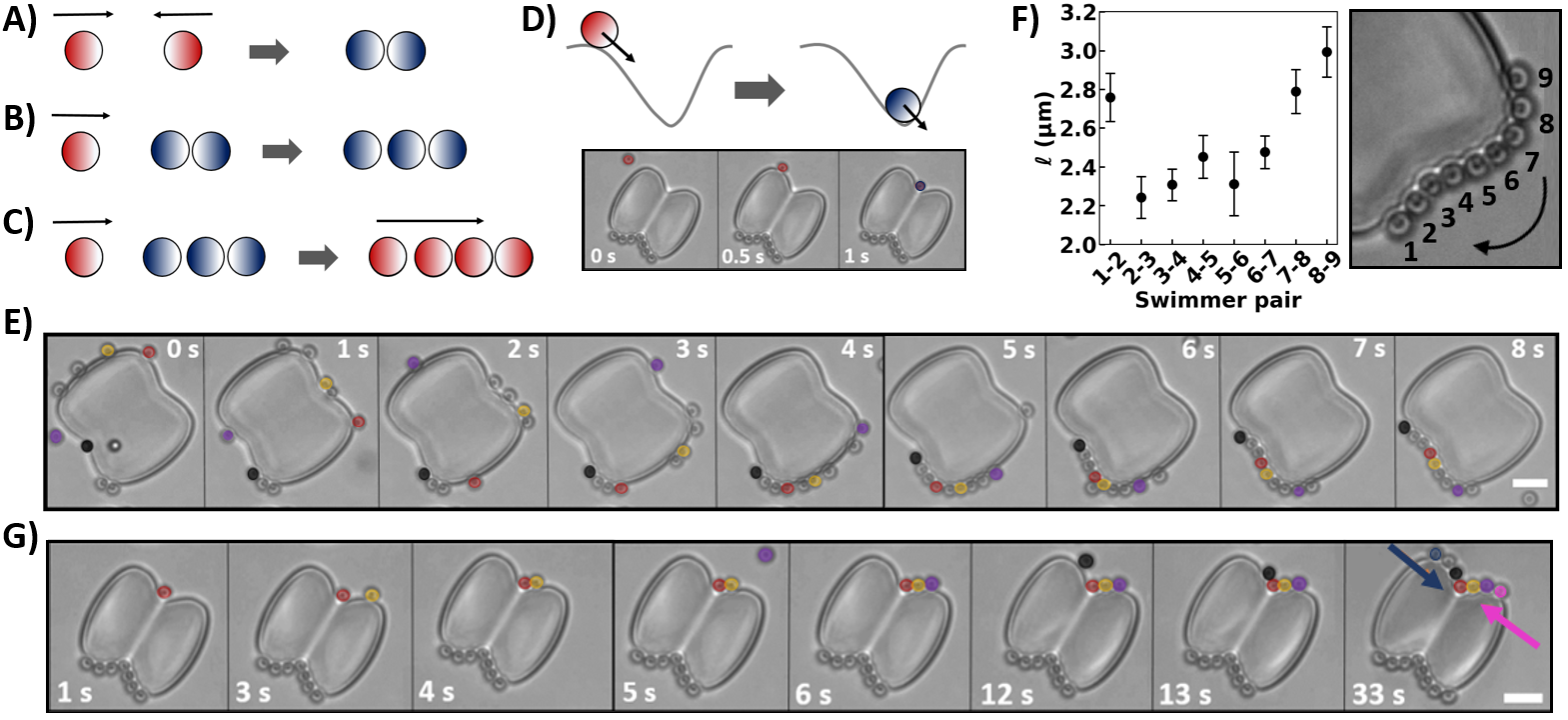}
    \caption{\textbf{Actively-assembled chains of microswimmers on peanut-shaped posts.}
    \textbf{(A-D)} Schematic representations of the conditions determining chain mobility. Red and blue colors indicate mobile and immobilized swimmers, respectively. The darker hemispheres represent the Pt side, which drives motion. \textbf{(A)} Two swimmers with opposing motion direction become immobile. \textbf{(B)} When a third swimmer approaches the cluster, it gets in close touch and also becomes stationary. \textbf{(C)} A fourth swimmer, entering from the same side as in (B), causes the chain to move again in the direction of the majority of swimmers. \textbf{(D)} In highly curved neck regions, swimmers stop temporarily because a reorientation is required, see also the light microscopy image at the bottom. \textbf{(E)} Dynamic chain formation, see also Supplementary Movie 5: initially, self-propelled swimmers headed by the red swimmer move clockwise in a train. After being stopped by an immobile cluster at the edge, the train becomes a compact chain that moves after further addition of swimmers. \textbf{(F)} Center-to-center distance between swimmer pairs within a mobile chain that self-propels in the direction of swimmer ``1", i.e. swimmer ``1" is at the front end while swimmer "9" is at the back end, see also the light microscopy image on the right. Swimmers at the ends maintain at all times larger distances from their neighbors, in comparison to those positioned in the middle. \textbf{(G)} Posts with a more negative neck curvature enhance the formation of compact and immobile chains, because the necks act as accumulation sites, see also Supplementary Movie 6. This time-series is a continuation of that in panel (D). Same color marks the same swimmer throughout each time-series in (E, G).}
    \label{fig:fig4}
\end{figure*}

The shape of the resulting effective potentials close to the minimum resembles a harmonic function, and hence we fit them with $y = \frac{1}{2}k(x-x_0)^2+y_0$ using a least-squares fit. This provides us with the depth of the potential wells $y_0$, the preferred distance $x_0$, and the interaction strength $k$, see SI for values. Interestingly, the confinement seems to weaken with increasing preferred distance of the swimmers, which would be in line with the intuitive picture of interacting solute gradients. However, the larger preferred swimmer separation and lower interaction strength might also be due to a difference in one of the participating swimmers.

Let us now turn to the origin of this activity-induced interaction. In view of our recent work~\cite{Ketzetzi2020,Ketzetzi2020sep}, we hypothesize that there is a short-ranged repulsion due to the pusher-type nature of the propulsion mechanism, and a long-ranged attraction due to osmotic flows along the substrate. To lowest order, the former interaction induces an \textit{outward} directed dipolar fluid flow along the symmetry axis of the swimmer~\cite{Campbell2019} that scales as $u_{\mathrm{dip}}(r,\theta) = \kappa \left( 3 \cos^{2} \theta - 1 \right) / ( 2 r^4 )$, while the latter induces an \textit{inward} directed flow due to osmosis along the substrate~\cite{Uspal2018} that scales as $u_{\mathrm{osm}}(r) = -\lambda/r^2$; the factors $\kappa$ and $\lambda$ indicate the respective strength of the outward and inward flows, $r$ the radial distance --- the decay accounts for the presence of a no-slip surface ---, and $\theta$ the angle with respect to the swimmer's orientation.

When the two contributions balance at a finite distance, comoving swimmers can assume a stable separation, see Figure~\ref{fig:fig3}F for a vector plot of the total velocity profile $u_{\mathrm{tot}}(r,\theta) = u_{\mathrm{osm}}(r) + u_{\mathrm{dip}}(r,\theta)$. The angular dependence shows a lemniscate zero-velocity contour, which does not account for the finite size of the swimmer. Clearly, our simple argument would allow for swimmer contact at a finite angle, which can be stabilized by imposing 1-dimensional motion. Figure 3G shows that when aligned head to tail, there is a separation $r_{0}$ that is stable, as indicated by the slope at the intercept. For swimmers comoving in the same direction we obtain a simple expression for the separation as a balance between the inward and outward flow strength: $r_{0} = \sqrt{\kappa / \lambda}$, see Figure 3H. Hence, this profile can indeed be recast into an effective potential, as in Figure~\ref{fig:fig3}D. The collective speed-up observed in Figure~\ref{fig:fig2}F may thus stem from long-ranged chemical gradients leading to an enhancement of the phoretic driving mechanism, or from a collective hydrodynamic effect, such as a drag reduction for the actively-assembled dimers and trimer. Lastly, note that according to this minimal model if one particle is immobilized, the second particle may approach much more closely, possibly even come into contact depending on the strengths of repulsion and self-propulsion; the latter should also hold for particles moving toward each other.

Intriguingly, we can test these predictions in our experiments, where we may be able to induce situations of close contact which would disturb the stable swimmer distance, by temporarily or even permanently stopping one of the swimmers. The latter we have already observed for immobilized pairs of swimmers, on the blue-colored posts in Figure~\ref{fig:fig1}A. These swimmers probably moved in opposite directions before their encounter immobilized them, see Figure~\ref{fig:fig4}A for a schematic drawing. More surprisingly, we also observed immobile clusters consisting of three particles, see Figure~\ref{fig:fig1}A. This implies that a third particle not only was able to approach two immobilized particles much closer than the previously observed 3~{\textmu}m minimum distance, in line with our model prediction, but also that it did not remobilize the cluster despite the uneven particle number and hence presumably unbalanced forces, see Figure~\ref{fig:fig4}B.  
 
To gain better control over the location and duration of the stops and test our hypotheses, we employed micro-printed posts with a peanut shape that feature regions of alternating positive and negative curvature. Swimmers passing through negative curvature points need time to reorient themselves to be able to move on, thereby enforcing temporary stops at precisely defined locations, see Figure~\ref{fig:fig4}D. The more curved the neck regions of these posts, the longer it takes for the swimmers to escape. In addition, these larger posts allow us to study the interactions and behavior of more than three particles.   

Equivalently to our observations on the circular posts, and previous work on channels~\cite{Das2015}, we verify that swimmers with opposing directions of motion hinder each other, see the bottom edge of the peanut-shaped post in Figure~\ref{fig:fig4}E ($t$=0~s). Similarly, a third swimmer is unable to disturb this configuration and joins the immobile cluster, see bottom edge of Figure~\ref{fig:fig4}E ($t$=1~s), where the black-colored swimmer joins the immobile dimer. Thus, such immobilization not only happens in the negative curvature region of the neck, but also along the positive curvature regions of the rounded peanut edges. A fourth swimmer joining the immobile trimer from the right leads to a balanced situation and an immobile tetramer cluster similar to Figure~\ref{fig:fig4}A, see the red-colored swimmer in Figure~\ref{fig:fig4}E at $t$=2~s and $t$=3~s. Based on our experiment we found that a fourth swimmer is able to remobilize an immobile trimer, whenever there are three particles pointing in the same direction, see Figure~\ref{fig:fig4}C. Generally in combining chains, we find remobilization whenever the number of swimmers pointing in one direction exceeds the number of swimmers pointing in the other direction by $\Delta n = 2$. Notice, however, that swimmers pointing in the direction opposite to the net motion do not need to reorient, they are simply pushed along. 

Based on the aforementioned simple conditions, we can furthermore deduce the directionalities of the swimmers in compact moving chains, either from their position or from the dynamic information of chain formation. For example, swimmers at the ends of long immobile chains always must have directions that point towards its center. Conversely, in this manner we can also predict the dynamics of a compact chain upon addition of a swimmer. 
 
The larger size of our peanut-shaped microprinted posts enables the attachment of multiple moving swimmers that can actively interact and dynamically self-assemble and disassemble. These posts therefore allow us to see how swimmers that move in trains along the post evolve into compact swimmer chains, see Figure~\ref{fig:fig4}E and Supplementary Movie 5. For example, four swimmers that move in the same clockwise direction form a train led by the red-colored swimmer in Figure~\ref{fig:fig4}E ($t$=0~s). Between $t$=0~s and $t$=1~s a fifth swimmer, initially swimming on the top of the post, enters the train of comoving swimmers from above, in front of the orange one. Interestingly, the swimmers following the red maintain stable distances while orbiting. That is, before being stopped by the immobilized cluster at the bottom-left corner, their in-between distances fluctuate around a preferred distance in accordance with our Figures~\ref{fig:fig2}~and~\ref{fig:fig3} findings. The immobile cluster acts as a stopping point that compacts the train of swimmers into a chain: one by one, the swimmers encounter and join the cluster in close contact, thereby creating a compact chain. Once the number of clockwise moving swimmers is at least by two greater than the number of counterclockwise moving swimmers in the immobile cluster, the entire chain sets into motion. Since the majority of swimmers are moving clockwise, the clockwise direction is imposed to the chain as a whole. 

After the chain is formed and remobilized, the distance between a swimmer and its neighbor depends on its position in the chain, see Figure~\ref{fig:fig4}F. Swimmers at the chain ends are further apart from their neighbors as opposed to the ones in the middle that are almost in touch. Swimmers at both ends are positioned at a center-to-center distance of $\approx$ (2.9 $\pm$ 0.2)~{\textmu}m from their neighbors, unlike the swimmers within the chain that move at distances of $\approx$ (2.4 $\pm$ 0.3)~{\textmu}m. For comparison, our swimmers have a diameter of (2.00 $\pm$ 0.05)~{\textmu}m, implying that indeed particles in the center of the chain are almost touching. Note that the distance of the swimmer pairs at the chain ends coincides with the minimum distance found for the swimmer pairs in the circular posts in Figure~\ref{fig:fig3}B. This observation further corroborates our hypothesis of a long-range attraction being present between the swimmers, which is balanced by a short-range repulsion. Because the attraction spans more than a single swimmer, the swimmers in the middle are more compacted than those at the end. 

Contrary to the dynamic collective motion we have just described, we furthermore find that highly curved necks capture swimmers for such long times, that in fact they act as semi-permanent stopping points, see Supplementary Movie 6. Although long chain formation is still possible, the self-assembled chains remain immobile due to the pinning of the swimmers by the neck, see Figure~\ref{fig:fig4}D and G: the motion of the red swimmer immediately stops at the neck at $t$=1~s. Thus, a highly curved neck not only prohibits collective motion, but also hinders single swimmer motion since it requires reorientation.

Besides an activity-induced self-assembly into compact chains, chains may also reorganize in time, see Supplementary Movie 7. In Figure~\ref{fig:fig5}A we follow a clockwise self-propelling chain consisting of ten swimmers. \begin{figure}[ht]
    \centering
     \includegraphics[width=1\linewidth]{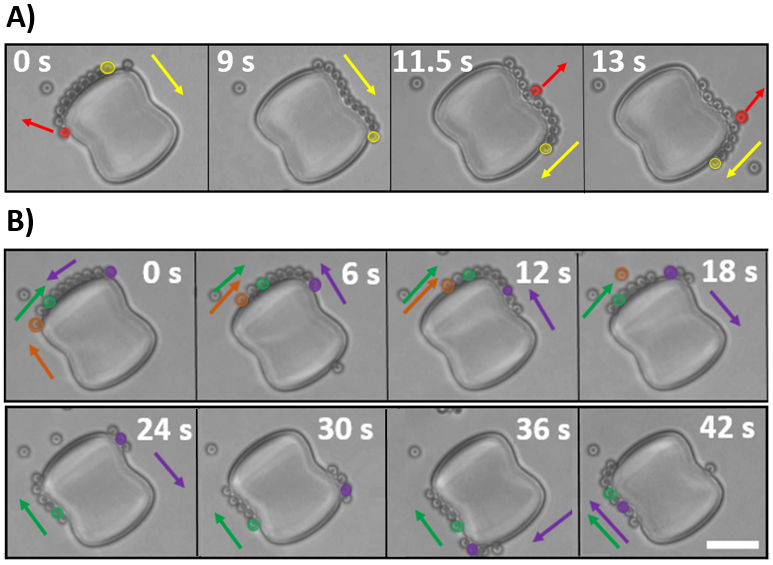}
    \caption{\textbf{Dynamics of actively-assembled chains: Effect of post topology on chain mobility.} Each color marks the same swimmer within each time-series, except the red color in (A) which marks three different swimmers all escaping the chain; corresponding arrows denote directionality. \textbf{(A)} When it reaches the rounded post edge, a swimmer escapes from the end of the self-propelled compact chain. Swimmers from the middle of the chain also escape during orbiting from the neck or the rounded post edge. See also Supplementary Movie 7. \textbf{(B)} Example of chain breakup: a chain self-propels in the clockwise direction and becomes immobile around the rounded post edge. Two swimmers with opposing direction are incorporated to the chain from the right, and the chain remobilizes in the opposite direction. When reaching the rounded edge, the chain end slows down, the chain breaks, and its end immobilizes. Upon arrival of new swimmers, the end remobilizes at the reverse direction eventually leading to the reformation of the entire chain at the opposite site of the neck. See also Supplementary Movie 8.}
    \label{fig:fig5} \vspace{-5pt}
\end{figure}During orbiting, swimmers may leave the chain in the following ways: (1) swimmers at the chain end may leave when they reach locations of comparatively high positive or negative curvature, in line with earlier findings for individual swimmers \cite{Das2015, Simcchen2015}, which here is also likely facilitated due to the larger distances to their neighbor. That is the case for the red swimmer at $t$=0~s, which leaves the chain when it reaches the rounded peanut edge (top left). (2) Likewise, swimmers from the middle may exit when they pass through locations where curvature varies. This situation is visible both at $t$=11.5~s and $t$=13.0~s in Figure~\ref{fig:fig5}A, where a center-chain swimmer highlighted in red escapes while passing through the negatively curved neck and positively curved corner, respectively. In both cases, the chain slowed down before the escape. We speculate that this is enhanced for swimmers with directionality that opposes the direction of motion of the chain.   

In addition to facilitating swimmer escape, locations with strong curvature variations furthermore enhance chain breakup and motion reversal, see Supplementary Movie 8 and Figure~\ref{fig:fig5}B as an example. A chain that moves clockwise slows down and becomes immobile when its middle is situated on the rounded edge. The swimmer from the middle leaves the chain while the entire chain remains stationary. At $t$=12~s two swimmers with the same counterclockwise direction as the purple swimmer enter the chain from the right. The entire chain gets remobilized and starts to move in the counterclockwise direction: when reaching the rounded edge, the end of the chain is being slowed down, then the chain breaks, leaving three swimmers immobilized on the rounded edge behind. The rest of the green-swimmer-led chain continues to move counterclockwise until it stops at the other neck side. From the stationary purple chain one swimmer escapes, and gets remobilized in the clockwise direction by two swimmers joining. When the purple chain encounters the immobile green chain, it lifts the immobilization and the entire chain continues in the clockwise direction, in line with what we observed previously in Figure~\ref{fig:fig4}E. 

\section*{Discussion}

In summary, catalytically self-propelled microswimmers show a plethora of striking collective effects in 1-dimensional environments. When moving in the same direction microswimmers ``cooperate". That is, they move at a greater speed the more particles join as well as at a preferred distance, larger than a minimum distance. This activity-induced interaction can be described by an effective potential of the order of few $k_BT$. Stopping the chain can overcome this minimum spacing of comoving swimmers, thereby leading to compact chains. These active chains show very rich dynamics themselves, with activity induced self-assembly, compaction, disassembly, breakup, and reformation. A simple rule seems to distinguish between mobility and immobility for these chains: mobility is achieved if the difference in the numbers of opposing swimmers is greater or equal to two. Once in contact, the notion of the potential well still persists, as distances within the self-assembled chains are not uniform, i.e. swimmers at the ends are less tightly packed. Curvature modulations of the posts allow spatial control over chain speed and reorganization, such as compaction, breakup, and immobilization, offering an exciting playground for testing efficient modes of transport. Overall, our findings unravel unprecedented behaviors of model microswimmers in 1-dimensional environments, thereby providing novel insights into activity-induced interactions. Insights into interactions and collective behavior of synthetic microswimmers are pivotal for applications that require increased swimming efficiency or directionality across different environments, as well as answering fundamental questions on the 1-dimensional phase behavior of active systems. 

\section*{Methods}

\paragraph*{{\normalfont \textbf{Particles.}}} Spherical latex particles based on polystyrene (2\% cross-linked) with diameter (2.00~$\pm$~0.05)~{\textmu}m, i.e. size polydispersity 2.5\%, were purchased from Sigma Aldrich. Pt-half-coated particles, with a Pt layer thickness of (4.7~$\pm$~0.2)~nm, were produced through physical vapor deposition, as described in Ref.~\cite{Ketzetzi2020,Ketzetzi2020sep}. 

\paragraph*{{\normalfont \textbf{3D printed structures.}}} Microstructures were produced with the commercially available microprinter Photonic Professional GT of Nanoscribe which uses two-photon lithography. The microprinter was equipped with a 63X oil-immersion objective (Zeiss, NA = 1.48) and used to print the 3D structures in oil mode. Microstructure designs were performed in Autodesk Inventor and processed with Describe. The microstructures were printed onto glass coverslips, pre-cleaned with isopropanol, using the commercial photoresist IP-L as a pre-polymer. After printing, the structures were developed by submersion in propylene glycol methylether acrylate for 15 min, followed by gently dipping into isopropanol three times to remove the unpolymerized photoresist. The structures were subsequently dried with gentle air flow. All procedure was done under yellow light. 

\paragraph*{{\normalfont \textbf{Imaging.}}} Pt-half-coated particles were dispersed in a 10\% aqueous H$_2$O$_2$ solution. Their motion was recorded above the planar walls with a ELWD 60x objective (S Plan Fluor, NA 0.7, zoomed at 1.5x i.e.~0.1 $\mu$m/px) mounted on an inverted Nikon Eclipse Ti microscope at a frame rate of 5 and 9 fps along the circular and peanut-shaped posts, respectively, within the first hour after sample preparation. 

\paragraph*{{\normalfont \textbf{Analysis.}}} Particle positions above the planar wall and along the circular posts were obtained using the Python tracking algorithm Trackpy~\cite{Trackpy}. The speed of all particles was determined using the time derivatives of spatial displacements at consecutive frames, see inset of Figure~\ref{fig:fig1}D for the speed distribution of a single particle measured in orbit for $\approx$ 4 min ($>$ 1200 frames). For swimmers on circular posts, the (arc) displacement was obtained according to Figure~\ref{fig:fig3}A. On the peanut-shaped posts, distances were obtained using the NIS-Elements Advance Research software package by Nikon. 

\paragraph*{{\normalfont \textbf{Acknowledgements}}}
D.J.K. gratefully acknowledges funding from the European Research Council (ERC) under the European Union's Horizon 2020 research and innovation program (grant agreement no. 758383). J.d.G. thanks NWO for funding through Start-Up Grant 740.018.013 and through association with the EU-FET project NANOPHLOW (766972) within Horizon 2020.

\clearpage
\bibliography{bibliography}

\end{document}